\documentclass[letterpaper, 10 pt, journal, twoside]{IEEEtran}

\usepackage{bm}
\usepackage{float}
\usepackage{amsmath}
\usepackage{amssymb}

\newcommand{\m}[1]{\pmb{#1}}
\newcommand{\norm}[1]{\left\lVert#1\right\rVert}
\newcommand{\abs}[1]{\left|#1\right|}
\newcommand{\sign}{\textrm{sign}}
\newcommand{\diag}{\textrm{diag}}

\newcommand{\T}{^\top}
\newcommand{\BM}{\begin{bmatrix}}
\newcommand{\EBM}{\end{bmatrix}}
\newcommand{\Rho}{\mathrm{P}}
\newcommand{\hM}{{ \m{\hat{M}} }}

\newcommand{\vsig}{{\m{\sigma}}}

\newcommand{\Rab}{{\m{R}^a_b}}
\newcommand{\Rba}{{\m{R}^b_a}}

\newcommand{\Rbd}{{\m{R}^b_d}}

\newcommand{\dRba}{{\m{\dot{R}}^b_a}}

\newcommand{\pba}{{\m{p}^b_a}}

\newcommand{\dpba}{{\m{\dot{p}}^b_a}}

\newcommand{\pbd}{{\m{p}^b_d}}

\newcommand{\kab}{{\m{k}^a_b}}

\newcommand{\nuaa}{{\m{\nu}^a_a}}
\newcommand{\omaa}{{\m{\omega}^a_a}}
\newcommand{\gaaa}{{\m{\gamma}^a_a}}

\newcommand{\nudd}{{\m{\nu}^d_d}}
\newcommand{\omdd}{{\m{\omega}^d_d}}
\newcommand{\gadd}{{\m{\gamma}^d_d}}

\newcommand{\dgaaa}{{\m{\dot{\gamma}}^a_a}}
\newcommand{\dgadd}{{\m{\dot{\gamma}}^d_d}}

\newcommand{\ea}{{\m{e}^a}}
\newcommand{\xea}{{{}^{x} e^a}}
\newcommand{\yea}{{{}^{y} e^a}}
\newcommand{\zea}{{{}^{z} e^a}}

\newcommand{\pej}{{{}^{p} e^j}}

\newcommand{\ej}{{\m{e}^j}}
\newcommand{\xej}{{{}^{x} e^j}}
\newcommand{\yej}{{{}^{y} e^j}}
\newcommand{\zej}{{{}^{z} e^j}}

\newcommand{\pe}{{{}^{p} e}}
\newcommand{\ke}{{{}^{k} e}}
\renewcommand{\oe}{{{}^{o} e}}

\newcommand{\mXi}{{\m{\Xi}}}

\newcommand{\dmXi}{{\m{\dot{\Xi}}}}
\newcommand{\mXid}{{\m{\Xi}^d\,}}
\newcommand{\dmXid}{{\m{\dot{\Xi}}^d\,}}

\newcommand{\iaa}{{\m{i}^a_a}}
\newcommand{\jaa}{{\m{j}^a_a}}

\newcommand{\iah}{{\m{i}^a_h}}
\newcommand{\jah}{{\m{j}^a_h}}

\newcommand{\pvsig}{{{}^p {\sigma}}}
\newcommand{\kvsig}{{{}^k {\sigma}}}
\newcommand{\ovsig}{{{}^o {\sigma}}}

\newcommand{\kdel}{{{}^k \delta}}

\newcommand{\ATAN}[2]{\mathrm{Atan2}\left({#1},{#2}\right)}

\ifCLASSINFOpdf
  \usepackage[pdftex]{graphicx}
\else
\fi

\begin{document}
\title{Robust 3D tracking control of an underactuated autonomous~airship}

\author{Wojciech Adamski\(^{1}\), Dariusz Pazderski\(^{1}\) and Przemysław Herman\(^{1}\)%
\thanks{Manuscript received: October 29, 2019; Revised: January 8, 2020; Accepted: April 27, 2020.}%
\thanks{This paper was recommended for publication by Jonathan Roberts upon evaluation of the Associate Editor and Reviewers' comments.
This work was supported by National Science Centre of Poland, as a research grant No.~2011/03/B/ST7/02524 and by the Poznań University of Technology grant No.~0211/SBAD/0911.} %
\thanks{\(^{1}\) Wojtek Adamski, Dariusz Pazderski and Przemysław Herman are with Institute of Automatic Control and Robotics, Poznań University of Technology (PUT), Piotrowo~3A, 60--965 Poznań, Poland 
        {\tt\footnotesize wojciech.adamski@put.poznan.pl}}%
\thanks{Digital Object Identifier (DOI): see top of this page.}
}%

\markboth{IEEE Robotics and Automation Letters. Preprint Version. Accepted April, 2020}
{Adamski \MakeLowercase{\textit{et al.}}: Robust 3D tracking control}

\maketitle

\begin{abstract}
    The paper  presents a new, robust control algorithm for position trajectory tracking in a 3D space, dedicated to underactuated airships. In order to take into account real  characteristics of such vehicles, and to reflect practically motivated constraints,  the algorithm assumes a highly uncertain system dynamics model.  The tracking problem is solved in a uniform way, without dividing it into subtasks considered  in 2D spaces,  thanks to the introduction of an auxiliary tracking error.

    The proposed controller is based on the sliding mode approach. 
    Its stability is investigated using Lyapunov theorem.
    Numerical simulations are conducted in order to verify properties of a closed-loop system for a generic model of the airship. 

    Performance of the control system is examined via experiments in various scenarios using a prototype airship. 
    The obtained results indicate that the control objectives are satisfied in practice with a reasonable accuracy. 
    Moreover, it is shown that the controller is robust to some bounded additive measurement perturbations and delays in the control loop. 
\end{abstract}


\IEEEpeerreviewmaketitle

\vspace{-0.05cm}
\section{Introduction}
\IEEEPARstart{A}{irships}, also called blimps,  are flying vehicles that take advantage of static buoyancy force for balancing weight.
For example, a 25-meter-long, 6-meter-wide ellipsoidal, helium-filled envelope at the sea level has a buoyancy of approximately 500 kg. It can lift a human pilot with the necessary equipment, or transport various types of apparatus or cargo. An autonomous airship with an electric propulsion system powered by solar panels, in theory, has a very long time of operation, which makes it an energy-efficient multi-purpose transport platform.

For most airships, the number of actuators is limited in order to optimize the buoyancy-weight ratio. Typically, only essential degrees of freedom of airships are actuated and the possibility of generating lateral thrust is nonexistent (or at least severely restricted). As a result, airships can be treated as underactuated mechanical systems, which are difficult to control.

The article proposes an original, experimentally verified
algorithm for trajectory tracking control for an underactuated airship. The main advantage of the described proposal is the application of a uniform design approach in the 3D space, which avoids the decomposition of control into subtasks in lower-dimensional spaces. The design also ensures low susceptibility to measurement errors and model uncertainties.

The synthesis of the control algorithm is performed in two main steps. In the first step, the so-called auxiliary error, which combines all components of the position error in $\mathbb{R}^3$ and a non-standard measure of orientation errors, is defined. This solution facilitates the design of the controller and makes it possible to deal with the underactuation issue. In the second step, the control rule which operates on the level of dynamics is designed. The rule guarantees the convergence of the auxiliary error to an arbitrarily small vicinity of zero.

There are some interesting articles which describe the problem of trajectory tracking or path following specific to airships. Some good examples are \cite{Azinheira2009} and \cite{Zheng2018a}, which discuss the backstepping technique and the adaptive sliding mode controller, respectively. Unfortunately, few articles include experimental results.

In the paper \cite{Kohno2005}, Kohno and Sasa describe the results of the implementation of an algorithm based on the principle of dividing the control task into regulation of orientation, altitude, and velocity.
In \cite{DePaiva2006a}, De Paiva et al. present experiments with controllers that use linearization techniques and are applied to separate tasks of controlling rotation and longitudinal motion on a plane.
Rao et al. in the paper \cite{Rao2007} use knowledge-based neural networks to control the angle of rotation around the vertical axis of the airship.
Yamada et al. in \cite{Yamada2007} describe the experimental results for an algorithm using linearization of dynamic equations of motion on a plane.
An interesting analysis of the topic of airship control is presented by Solaque and Lacroix in \cite{Solaque2007}. They compare the performance of a PID controller, Generalized Predictive Control, and a controller which uses extended linearization. However, all tested solutions are implemented separately for each degree of freedom.
In \cite{Fukao2008}, Fukao et al. present the operation of three combined algorithms performing tracking a straight line on a plane, a turning maneuver and altitude regulation.
Another experiment using fuzzy logic to control the angle of rotation around the vertical axis of an object is described in \cite{Dai2011}.
Saiki et al. in \cite{Saiki2011} present the results of experiments with the use of a PID controller to regulate the altitude and angle of attack, and the optimal motion controller on a plane.

A cascading approach to control, dividing the problem into the task of controlling orientation and the task of controlling position using the method of linearization of trajectory tracking error is presented in \cite{Zheng2013}, together with experimental verification.
Adding to the cascade a high-level controller for generating a specified trajectory to solve the problem of underactuation is discussed in work \cite{Liesk2013}, which describes in detail the speed and orientation controller using back-stepping and square optimization.
The results of the ``Bang-Bang'' controller based on the analysis of the phase plane of a simplified model of dynamics in one degree of freedom are presented in \cite{Miao2016}. Wang et al. in \cite{Wang2018} present an algorithm of altitude control which uses state observers. They take into account the delay in the control signal path and relatively small disturbances.
Of particular interest are the papers \cite{Fedorenko2016} and \cite{Bechlioulis2017}, the first of which presents experimental results of the implementation of the algorithm described in \cite{Pshikhopov2010} (which proposes a robust controller using the inverse dynamics method). The disadvantage of the algorithm is, however, the fact that it requires the knowledge of both external disturbances and those resulting from the dynamics of the object. The experiment presented in the paper is about tracking the trajectory on a plane in the absence of significant external disturbances.

Bechlioulis et al. in \cite{Bechlioulis2017} propose a model-free control algorithm. The authors consider its implementation for "torpedo-like" and "unicycle-like" objects. This work contains a very promising solution, but
the authors do not analyze the robustness of the algorithm
with respect to control parameters, measurement errors, or the presence of significant external disturbances.

Michałek et al. in \cite{Michaek2019}
describe a VFO-ADRC cascaded trajectory tracking control system designed for rigid-body vehicles moving in the 3D space with non-banked maneuvering.
The performance of the system strongly depends on the quality of measurements. The results are confirmed by experiments.

There are also papers documenting experimentally verified control algorithms for surface and submarine ships. Experimental results for an underactuated submarine are presented in \cite{Antonelli2001}, in which the authors propose a PD controller extended by an adaptive noise compensation mechanism. In \cite{Lefeber2003}, a non-linear cascade controller is verified for an underactuated surface ship.
In \cite{Soylu2016}, the authors present a complete control system based on an adaptive version of the PID controller, used in a submarine, where the mathematical model of motion does not include the degrees of freedom associated with rotation around the \(x\)- and \(y\)-axes.
Martin and Whitcomb in \cite{Martin2018} compare the results of experiments with a PD controller and two different model-based controllers. The experiments were performed for a fully actuated underwater object moving in the 3D space. Although the model-based drivers considered in the study yielded better results, the authors emphasize their high sensitivity to inaccuracy in estimating dynamic parameters.

\begin{figure}[t]
    \centering
    \includegraphics[width=0.7\linewidth]{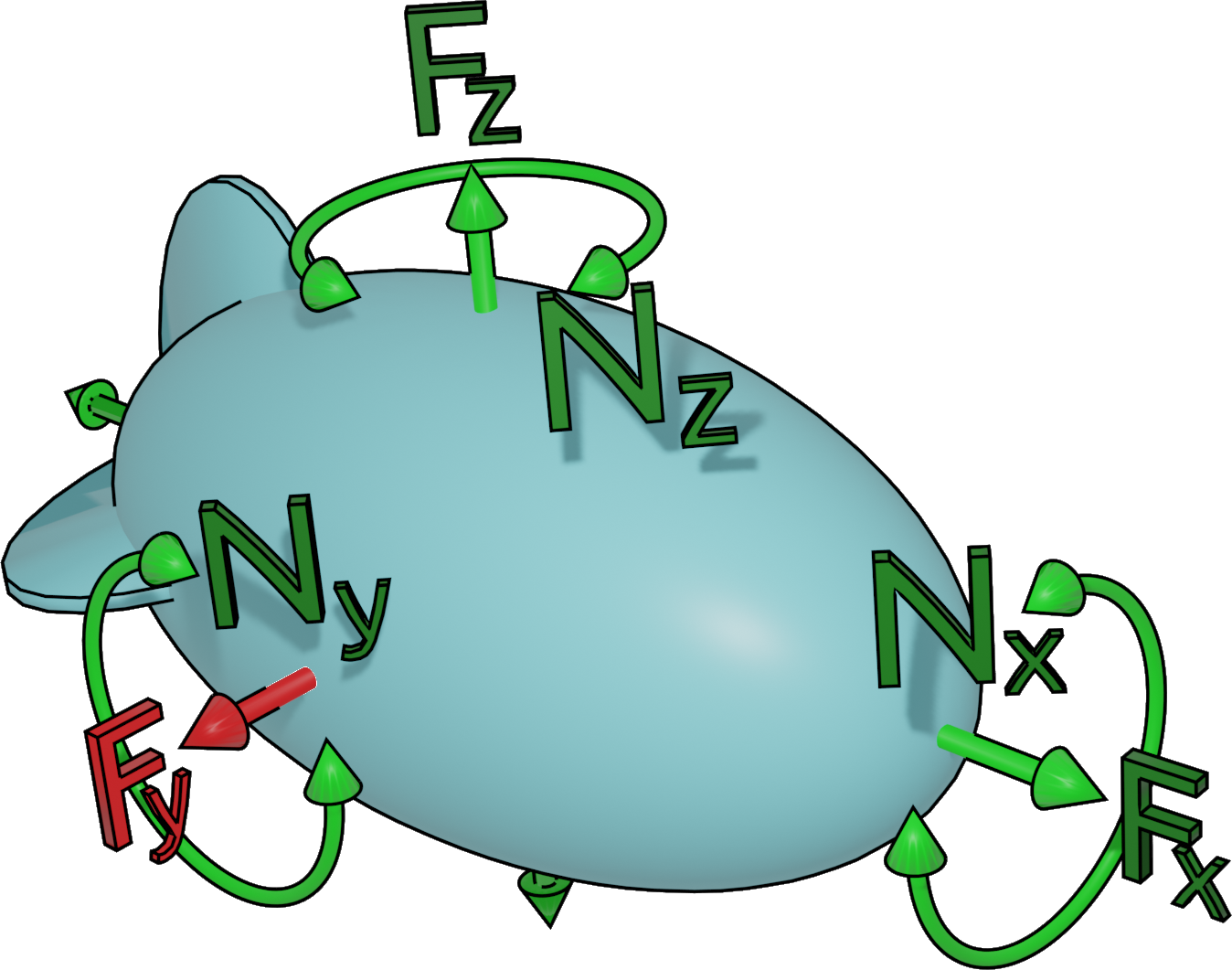}
    \vspace{-0.5em}
    \caption{The visualization of an airship with indicated directions of forces and torques \(\tau=\left[ F_x \, F_y \, F_z \, N_x \, N_y \, N_z \right]\T\). In~the direction marked in red, there is no control signal.}
    \vspace{-1.5em}
    \label{fig:airship_vis}
\end{figure}

In contrast to the solutions discussed above, the approach presented in this paper treats the problem of object motion in the 3D space as a whole, i.e. the task of motion control is not divided into subtasks such as control of motion on a plane, altitude control, orientation control or speed control.

Motion control is provided by a controller operating on the level of
dynamics and inspired by the sliding mode approach,
characterized by relatively low sensitivity to disturbances,
parametric and structural uncertainties, and measurement errors.

To achieve a uniform solution and to alleviate the underactuation
problem (cf. Fig 1), previously mentioned auxiliary errors are used.
It should be emphasized that the design of the controller proposed in
this paper takes into account some implementation constraints which
can be met in practice. Hence, the structure of the controller is
chosen so as to find a proper balance between theoretical and
experimental contributions.
The performance of the algorithm was verified both by simulation and
experimentation in the presence of disturbances and measurement errors
in the task of tracking a prescribed 3D trajectory.

The paper is organized as follows. Section \ref{sec:prereq_ctrl_obj} provides the description of the object and control objectives. Section \ref{sec:ctrl_law} explains the algorithm and Sections \ref{sec:num_valid} and \ref{sec:exp_valid} provide results of numerical and experimental verification, respectively. Section \ref{sec:conclusions} contains conclusions.

\section{PREREQUISITES AND CONTROL OBJECTIVES} \label{sec:prereq_ctrl_obj}     
\subsection{Notation}

\vspace{-0.5em}
\begin{table}[!h]
	\renewcommand*{\arraystretch}{1.3}
	\centering
	\begin{tabular}{p{1.6cm}p{6.4cm}}
		\(\pba\)                                                                   & vector of position of the origin of frame $a$ in frame $b$                                                             \\ \hline
		$\Rab\in\mathrm{SO}(3)$                                                    & rotation matrix which defines the transformation from frame $a$ to frame $b$                                           \\ \hline
		$\mathbb{R}_{>0}^n$                                                        & set of vectors in $\mathbb{R}^n$ with all positive components                                                          \\ \hline
		$\mathbb{R}_{\geqslant 0}^n$                                               & set of vectors in $\mathbb{R}^n$ with all non-negative components                                                      \\ \hline
		$\mathcal{E}$                                                              & $\mathbb{R}_{\geqslant 0}\times[0,\sqrt{2}]\times[0,\sqrt{2}]$                                                         \\ \hline
		$\mathcal{E}_0$                                                            & $\mathbb{R}_{>0}\times(0,1)\times(0,1)$                                                                                \\ \hline
		$\diag([\alpha_1 \makebox[1em][c]{,\hfil.\hfil.\hfil.\hfil,}\, \alpha_n])$ & diagonal matrix of size $n\times n$ with diagonal entries given by real numbers $\alpha_1, \alpha_2, \ldots, \alpha_n$ \\ \hline
		$\m{1}$                                                                    & $\diag([1 \, 1\, 1\, 1\, 1\, 1])$                                                                                      \\ \hline
		$\m{S} \left( \cdot\right) \in \mathrm{so}(3)$                          & skew-symmetric matrix                                                                                                  \\ \hline
		$\underline{\lambda}\left\{\m{X}\right\}$                                  & the smallest eigenvalue of matrix \(\m{X}\)
	\end{tabular}
\end{table}
\vspace{-0.8cm}
\subsection{Airship model}

\begin{figure}[th]
	\centering
	\includegraphics[]{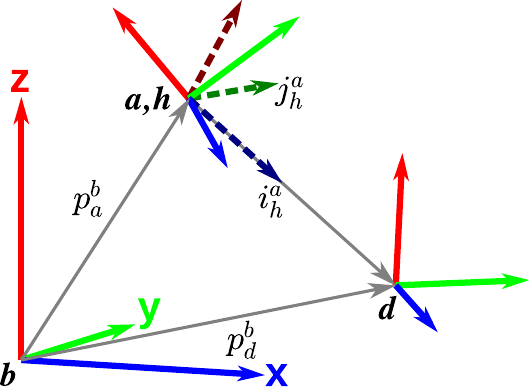}
	\vspace{-0.5em}
	\caption{Visualization of reference frames:
		\(b\) -- base frame (inertial),
		\(a\) -- body frame (in the center of mass),
		\(d\) -- trajectory frame (reference goal),
		\(h\) -- auxiliary frame (dashed line),
		\(\iah\) -- unit vector parallel to the line segment connecting the origins of frames \(a\) and \(d\),
		\(\jah\) -- unit vector orthogonal to \(\iah\) and parallel to the global plane \(XY\).}
	\vspace{-1.5em}
	\label{fig:frames}
\end{figure}

In order to construct a model of the blimp we introduce the following reference frames: base frame $b$ (inertial) and body frame $a$ attached at the center of mass of the blimp, cf. Fig. \ref{fig:frames}. Next, we recall the following kinematics in the 3D space \cite{BestaouiSebbane2011}:
\begin{align}
	\label{eq:kinematyka_01}
	\dpba = & \Rba \nuaa,                     \\
	\label{eq:kinematyka_02}
	\dRba = & \Rba \m{S}\left( \omaa \right),
\end{align}
where \(\nuaa\in\mathbb{R}^3\) and \(\omaa\in\mathbb{R}^3\) stand for linear and angular velocities, respectively, expressed in the body frame.
Denoting these velocities by \(\gaaa = \left[ \nuaa^\top\ \omaa^\top\right]^\top\in\mathbb{R}^6\), we can consider the following dynamics \cite{Li2011}:
\begin{align}
	\label{eq:dynamika}
	\dgaaa = \m{M}^{-1} \left(\m{\tau} - \m{f} \left( \gaaa, \Rba \right)\right),
\end{align}
where \(\m{M}\in\mathbb{R}^{6\times 6}\) is the inertia matrix, \(\m{\tau}\in\mathbb{R}^6\) is the force/torque control input and \(\m{f}\in\mathbb{R}^6\) stands for other dynamics terms, defined by
\begin{align}
	\label{eq:dynamika_details}
	\m{f} \left( \gaaa, \Rba \right) = \m{C}\left(\gaaa\right) \gaaa + \m{D}\left(\gaaa\right) + \m{G}\left(\Rba\right),
\end{align}
while \(\m{C}\left(\gaaa\right)\gaaa\) describes Coriolis and centrifugal forces, \(\m{D}\left(\gaaa\right)\) denotes damping forces, \(\m{G}\left(\Rba\right)\) combines the gravity and boyancy terms.
A detailed formula for $\m{f}$ for airships can be found in \cite{Li2011}; however, due to the properties of the control algorithm proposed in the paper, we do not consider this formula in depth here.
From now on, the deficiency in control is assumed, namely, the second component of \(\m{\tau}\) which indicates the lateral thrust $F_y$, cf. Fig.~\ref{fig:airship_vis}, is zero. Thus, the control system described by  \eqref{eq:kinematyka_01}, \eqref{eq:kinematyka_02} and \eqref{eq:dynamika} is underactuated.
\subsection{Control objectives}

Let us consider the trajectory frame \(d\) depicted in Fig.~\ref{fig:frames} with its position in the base frame given by  \(\m{p}_d^b\in\mathbb{R}^3\), and let us define the following tracking error in frame \(a\):
\begin{align}
	\ea= \left[\xea\ \yea\ \zea\right]^\top := \Rab \left(\pbd - \pba\right)\in\mathbb{R}^3.
\end{align}

The investigated control problem can be stated as follows. Assuming that the dynamics of an underactuated blimp is not fully known, find input \(\m{\tau}\) such that for the reference position trajectory \(\m{p}_d^b\) which satisfies \(\norm{\m{\dot{p}}_d^b} < \infty\) and \(\norm{\m{\ddot{p}}_d^b} < \infty\), the tracking error \(\ea\) is bounded and converges to a certain neighborhood of zero in the presence of bounded disturbances.

\section{CONTROL LAW DESIGN} \label{sec:ctrl_law}
It should be emphasized that since the second entry of $\m{\tau}$ is zero, the lateral tracking error component $\yea$ cannot be controlled trivially. Owing to this fact, the authors
propose to employ an auxiliary tracking error $\m{e}$ by using an Euclidean norm of the position error and specific measures of orientation errors, similarly as it can be done for planar noholonomic kinematics \cite{PSK:2012}, and torpedo-like vehicles, cf. \cite{Michaek2019}. Due to this approach, the control input \(\m{\tau}\) enables an independent action on each component of \(\m{e}\) while simultaneously guaranteeing that for bounded \(\m{e}\) the original tracking error \(\ea\) is also bounded.

\subsection{Auxiliary errors}
\begin{figure}[t]
	\centering
	\includegraphics[]{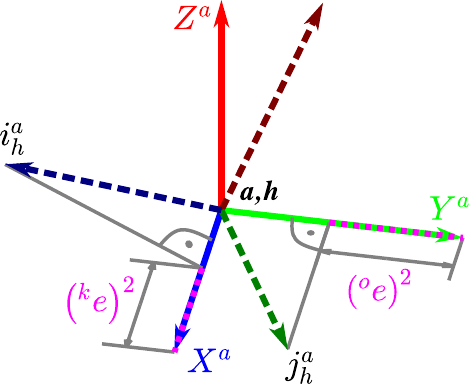}
	\vspace{-0.5em}
	\caption{Geometric interpretation of auxiliary errors \(\ke\) and \(\oe\). Solid unit vectors denote body frame \(a\), and dashed unit vectors are axes of auxiliary frame \(h\).}
	\vspace{-0.5cm}
	\label{fig:err_pe_ke_oe}
\end{figure}

In order to explain in detail the concept of the auxiliary error, let us consider the auxiliary frame \(h\) (dashed lines) in Figs.~\ref{fig:frames} and \ref{fig:err_pe_ke_oe}. Unit vectors \(\iah\) and \(\jah\) define $x$- and $y$-axes of the auxiliary frame, respectively. Note that vector \(\iah\) is parallel to the line segment connecting the origins of frames \(a\) and \(d\). Namely, it starts at the origin of the body frame \(a\) and points to the origin of the trajectory frame \(d\). The unit vector \(\jah\) is orthogonal to \(\iah\) and parallel to the global plane \(XY\). The latter is a design assumption resulting from practical reasons.

Figure \ref{fig:err_pe_ke_oe} shows only body (solid unit vectors) and auxiliary (dashed unit vectors) frames. The projections of \(\iah\) on the axis \(X^a\) and \(\jah\) on the axis \(Y^a\) determine the lengths of the dashed purple line segments, which are geometric interpretations of \(\left(\ke\right)^2\) and \(\left(\oe\right)^2\), the squares of auxiliary errors dependent on the orientation of the vehicle. To describe them algebraically,
let us define errors \(\m{e}\in\mathcal{E}\), \(\ej\in\mathbb{R}^3_{\geqslant 0} \) and \(\pej\in\mathbb{R}_{\geqslant 0}\):
\begin{align}
	\m{e} = \BM \pe \\ \ke \\ \oe \EBM, \, \ej = \m{S}\left( \kab \right) \ea = \BM \xej \\ \yej \\ \zej \EBM, \, \pej = \norm{\ej},
\end{align}
where \(\kab\) is the third column of the rotation matrix \(\Rab\), and
\begin{align}
	\pe = \norm{\ea},      \,
	\ke = \sqrt{1-\frac{\xea}{\pe}},  \,
	\oe = \sqrt{1-\frac{\yej}{\pej}}.
	\label{eq:ke}
\end{align}

\subsection{Design of the controller}
To ensure the robustness of the controller to unknown dynamics, the sliding control approach is employed. Taking into account error $\m{e}$ and its time derivative $\dot{\m{e}}$, we define the following sliding hyper-plane:
\begin{align}
	\label{eq:sigma}
	\vsig = & \m{\dot{e}}+\m{K} \left( \m{e} -\m{\delta}\right),
\end{align}
where $\m{\sigma}\in\mathbb{R}^3$ is the sliding variable, $\m{\delta} = \left[{}^p \delta\ {}^k \delta\ {}^o \delta \right]\T\in\mathcal{E}_0$ is the offset, and $\m{K} = \diag\left(\left[ k_p\ k_k\ k_o\right]\right)\succ 0$ denotes the gain matrix. It can be easily shown that for $\m{\sigma}=\m{0}$
\begin{equation}
	\lim_{t\rightarrow\infty} \m{e}(t)=\m{\delta},
\end{equation}
namely, that the error trajectory converges to $\m{\delta}$, which is selected at some non-zero distance from the origin.
The introduction of parameter \(\m{\delta}\) is essential due to stability conditions (\ref{eq:stability_conditions_on_errors}), which will be explained in Section \ref{sec:stability_analysis}. It is important to note that guaranteeing that $\m{e}\in \bar{\mathcal{E}}$, where $\bar{\mathcal{E}} = \left\{\m{e}\in\mathcal{E}: {}^pe<\infty\right\}$, allows us to ensure that the tracking error $\m{e}^a$ is bounded. Taking  into account \eqref{eq:ke} and making simple calculations, we can derive the following:
\begin{equation}
	\label{eq:wlasnosc_ke}
	\left(\yea\right)^2 = \left(\left(1-\left({}^ke\right)^2\right)^{-2} - 1\right)\left(\xea\right)^2-\left(\zea\right)^2.
\end{equation}

\begin{figure}[t]
	\centering
	\includegraphics[width=0.6\linewidth]{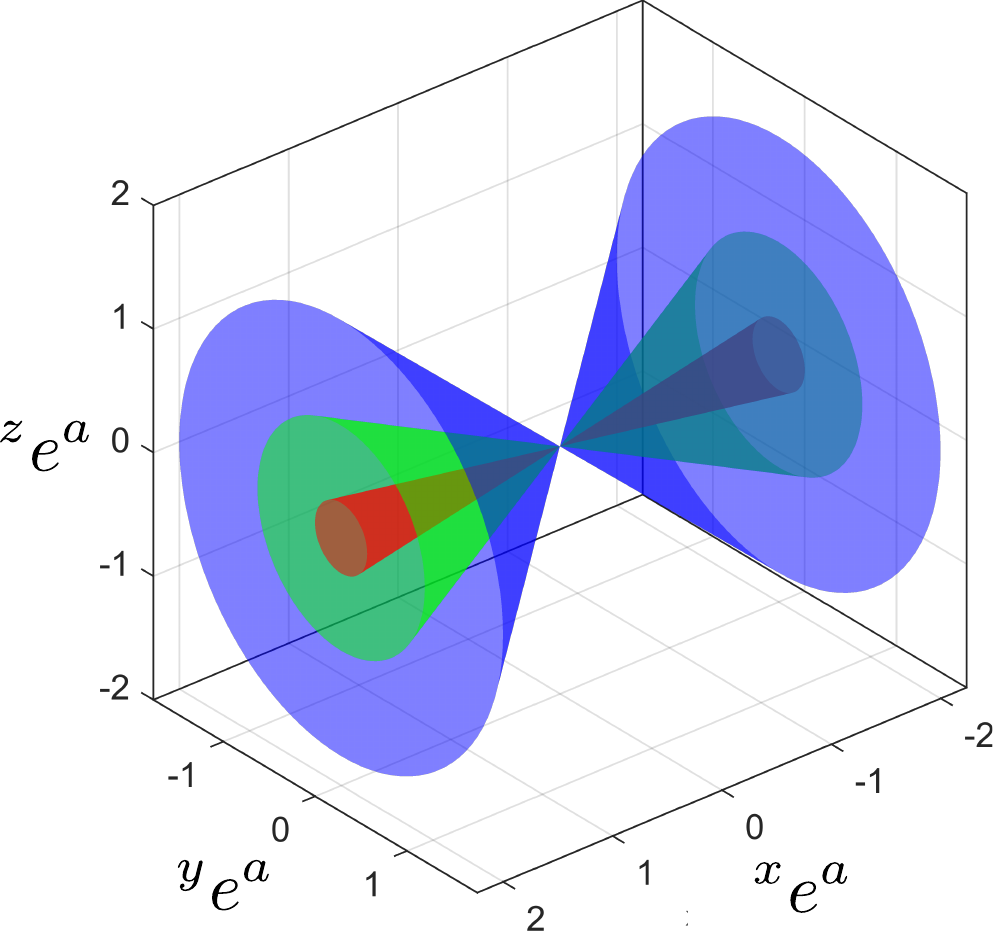}
	\vspace{-0.3cm}
	\caption{Error cone. The three cases are obtained for: ${}^ke= 0.1$ (red), ${}^ke= 0.3$ (green) and ${}^ke=0.5$ (blue).}
	\vspace{-0.4cm}
	\label{fig:err_cone}
\end{figure}

From \eqref{eq:wlasnosc_ke} it follows that $\ea$ lies on a double cone with the vertex at the origin (cf. Fig\ref{fig:err_cone}), which is parameterized by ${}^ke$ (for ${}^ke=1$ this cone degenerates to the $y$-$z$ plane). Since $\ke\rightarrow \kdel \in (0,1)$ we can see that the lateral position error $\yea$ is bounded and decreases with the decrease of $\kdel$ (recall that ${}^pe$ is assumed to be bounded). As a result, the lateral control can be substitued by stabilisation of $\ke$ at $\kdel$, which can be achieved in a more convienient way. This property is achieved thanks to the proposed form of the auxiliary errors.

In order to facilitate the design of the controller, we express $\dot{\m{e}}$ as follows:
\vspace{-0.1cm}
\begin{align}
	\label{eq:dot_e}
	\m{\dot{e}} = \mXi \gaaa + \mXid \gadd,
\end{align}
\vspace{-0.1cm}
where:
\vspace{-0.1cm}
\begin{small}
	\begin{align}
		\gadd = \BM \nudd \\ \omdd \EBM,\,\iah = \frac{\ea}{\pe}, \, \jah = \frac{\ej}{\pej}, \, \iaa = \BM 1 \\ 0 \\ 0\EBM, \, \jaa = \BM 0 \\ 1 \\ 0 \EBM,
	\end{align}
\end{small}
\vspace{-0.2cm}
\begin{small}
	\begin{align}
		\mXi =                                                                    & \diag\left(\BM \frac{1}{\pe}                                       & \frac{1}{2 \pe \ke} & \frac{1}{2 \pej \oe}\EBM\right) \nonumber \\
		\label{eq:mXi}
		\times                                                                    & \BM
		- \ea\T                                                                   & \m{0}                                                                                                                      \\
		- \left( \frac{\xea}{\pe} \iah\T \!\! - \iaa\T \right)                    & - \left( \iaa\T \m{S}(\ea) \right)                                                                                         \\
		- \! \left( \frac{\yej}{\pej} \jah\T \!\! - \jaa\T \! \right) \m{S}(\kab) & \left( \frac{\yej}{\pej} \jah\T\!\! - \jaa\T \! \right) \m{S}(\ej)                                                         \\
		\EBM.
	\end{align}
\end{small}
The form of \(\mXi\) and \(\mXid\) is the result of factorization of \(\m{\dot{e}}\) with respect to \(\gaaa\) and \(\gadd\). In the assumed
approach, the impact of \(\mXid\)-dependent components is considered as a disturbance. In order to maintain the readability of the paper, the detailed expression of \(\mXid\) is omitted.

In order to establish the stability of the sliding variable $\m{\sigma}$ in a certain neighborhood of zero, we propose the following control law:
\vspace{-0.1cm}
\begin{align}
	\label{eq:tau}
	\m{\tau} = & - \hM \m{\Rho} \mXi\T \sign{(\vsig)},
\end{align}
\vspace{-0.1cm}
where:
\begin{align}
	\label{eq:sign_sigma}
	\sign{(\vsig)} =                    &
	\BM \frac{\pvsig}{\abs{\pvsig}}                                                                            \\ \frac{\kvsig}{\abs{\kvsig}} \\ \frac{\ovsig}{\abs{\ovsig}}\EBM
	= \BM \frac{1}{\abs{\pvsig}}        & 0                                      & 0
	\\ 0 & \frac{1}{\abs{\kvsig}} & 0
	\\ 0 & 0 & \frac{1}{\abs{\ovsig}} \EBM
	\BM \pvsig                                                                                                 \\ \kvsig \\ \ovsig \EBM \\
	=                                   & \diag\left(\vsig^*\right) \vsig, \quad
	\vsig^* =\BM \frac{1}{\abs{\pvsig}} & \frac{1}{\abs{\kvsig}}                 & \frac{1}{\abs{\ovsig}} \EBM %
\end{align}
and $\m{\Rho} = \diag \left( \left[ \rho_u \, 0 \, \rho_w \, \rho_p \, \rho_q \, \rho_r \right] \right)$, $\rho_u,\, \rho_w,\, \rho_p,\, \rho_q,\, \rho_r  >0$ is an auxiliary gain matrix and \(\hM\in\mathbb{R}^{6\times 6}\) denotes the estimated mass matrix.

Matrix \(\hM\) must satisfy two conditions. The first one is due to stability requirements and can be stated as:
\begin{align}
	\label{eq:cond_Mest}
	\lambda_{e}=\underline{\lambda}\left\{\m{M}^{-1} \hM - \m{1} \right\} \geq 0.
\end{align}
The second condition is necessary to ensure that the second element of \(\m{\tau}\) is equal to zero. The condition results from the control rule (\ref{eq:tau}) and states that the second row of matrix \(\hM\) takes the following form: \(\left[0\,\beta\,0\,0\,0\,0\right],\, \beta\in\mathbb{R}\).

\vspace{-0.3cm}
\subsection{Stability analysis}\label{sec:stability_analysis}
Let us define the following Lyapunov-like function: $V := \frac{1}{2} \norm{\vsig}^2$. To ensure the convergence of the closed-system trajectory in a finite time to the sliding hyper-plane \(\m{\sigma}=\m{0}\), the following condition is assumed:
\begin{align}
	\label{eq:warunek_dotV_01}
	\dot{V} & \leq -\mu \norm{\vsig}, \quad \mu > 0.
\end{align}
Using dynamics (\ref{eq:dynamika}) and error derivative (\ref{eq:dot_e}), we can derive the following:
\begin{small}
	\begin{align}
		\dot{V} = & \vsig\T \m{\dot{\sigma}}
		\stackrel{(\ref{eq:sigma})}{=} \vsig\T \left(\m{\ddot{e}}+\m{K}\m{\dot{e}}\right)	\stackrel{(\ref{eq:dynamika}, \ref{eq:dot_e})}{=}\vsig\T \mXi \m{M}^{-1}
		\nonumber                                                                                                                                           \\
		+         & \vsig\T \left(-\mXi \m{f}\left(\gaaa,\Rba \right)+\dmXi\gaaa + \mXid \dgadd+ \dmXid \gadd+\m{K} \m{\dot{e}}\right)\label{eq:dV_stage1}.
	\end{align}
\end{small}
Next, we treat the terms in the last bracket of \eqref{eq:dV_stage1} as a disturbance and we define
\begin{align}
	\m{g} & =\m{g}\left(\pba, \Rba,\pbd, \Rbd,\gaaa, \gadd, \dgadd \right)\nonumber \\ &=-\mXi \m{f}\left(\gaaa,\Rba \right)+\dmXi\gaaa + \mXid \dgadd+ \dmXid \gadd+\m{K} \m{\dot{e}}.\label{eq:control:g_disturbance}
\end{align}
Consequently, \eqref{eq:dV_stage1} can be written as: $\dot V = \vsig\T \mXi \m{M}^{-1}\m{\tau} + \vsig\T \m{g}$, and after applying the control law \eqref{eq:tau}, the derivative $\dot{V}$ becomes:
\begin{align}
	\dot{V} = & - \vsig\T \mXi \m{M}^{-1} \hM \m{\Rho} \mXi\T \sign{(\vsig)} + \vsig\T \m{g}
	\nonumber                                                                                                                     \\
	=         & - \vsig\T \mXi \left( \m{M}^{-1} - \hM^{-1} + \hM^{-1} \right) \hM \m{\Rho} \mXi\T \sign{(\vsig)} + \vsig\T \m{g}
	\nonumber                                                                                                                     \\
	=         & - \vsig\T \mXi \left( \m{M}^{-1} \hM - \m{1} \right)  \m{\Rho} \mXi\T \sign{(\vsig)}
	\nonumber                                                                                                                     \\
	          & - \vsig\T \mXi \m{\Rho} \mXi\T \sign{(\vsig)}
	+ \vsig\T \m{g}.
\end{align}
Assuming that condition (\ref{eq:cond_Mest}) is satisfied, inequality (\ref{eq:warunek_dotV_01}) holds if:
\vspace{-0.2cm}
\begin{align}
	\label{eq:warunek_dotV_02}
	- \vsig\T \mXi \m{\Rho} \mXi\T \diag\left(\vsig^*\right) \vsig
	+ \vsig\T \m{g}
	 & \leq -\mu \norm{\vsig}.
\end{align}
In order to investigate \eqref{eq:warunek_dotV_02} more thoroughly, we take advantage of the following decomposition of matrix \(\m{\Rho}\):
\begin{align}
	\m{\Rho}=\m{\hat{\Rho}} \cdot \m{\hat{1}},
\end{align}
where \(\m{\hat{\Rho}} = \diag\left(\left[\rho_u\ \hat{\rho}\ \rho_w \ \rho_p\ \rho_q\ \rho_r \right]\right)\), \(\hat{\rho} > 0\) and
\(\m{\hat{1}}~=~\diag{\left(\left[1\ 0\ 1\ 1\ 1\ 1\right]\right)}\).

Next, we introduce the following smallest eigenvalues $\lambda_0=\underline{\lambda}\left\{\mXi \, \m{\hat{1}} \, \mXi\T\right\}$, $\lambda_\rho=\underline{\lambda}\left\{\m{\hat{\Rho}}\right\}$ and \(\lambda_\sigma = \underline{\lambda}\left\{ \diag\left(\vsig^*\right)\right\}=\frac{1}{\abs{\bar{\sigma}}}\), \(\abs{\bar{\sigma}} \triangleq \max\left(\abs{\pvsig}, \abs{\kvsig}, \abs{\ovsig} \right)\).
As a result, we can rewrite condition (\ref{eq:warunek_dotV_02}) as follows:
\begin{align}
	- \vsig\T \mXi \m{\Rho} \mXi\T \diag\left(\vsig^*\right) \vsig
	+ \vsig\T \m{g}
	\leq
	- \frac{\lambda_\rho \lambda_{0}}{\abs{\bar{\sigma}}} \vsig\T \vsig
	+ \vsig\T \m{g}
	\nonumber \\
	\leq
	- \frac{\lambda_\rho \lambda_{0}}{\abs{\bar{\sigma}}} \norm{\vsig}^2
	+ \norm{\vsig} \norm{\m{g}}
	\leq
	-\mu \norm{\vsig}.\label{eq:stability:cond1}
\end{align}
Taking into account that $\inf \left( \frac{\norm{\vsig}}{\abs{\bar{\sigma}}}  \right) = 1$, we can easily conclude that \eqref{eq:stability:cond1} is satisfied for
\begin{align}
	\label{eq:warunek_dotV_04}
	\lambda_{\rho} & \geq \frac{\mu+\norm{\m{g}}}{\lambda_0}.
\end{align}
Hence, for any bounded \(\norm{\m{g}}\) and positive \(\lambda_0\), there exists a finite \(\lambda_\rho\) and condition (\ref{eq:warunek_dotV_01}) holds. It implies that if $\lambda_0>0$, we can select finite gains in $\m{P}$. Conversely, for $\lambda_0=0$ there is a singularity. However, it can be proved that \(\lambda_0\) is equal to zero only when the trajectory $\m{e^a}(t)$ lies on one of the following three lines in the 3D:
\begin{align}
	\label{eq:war_zero_lambda_0_01}
	\xea = 0, & \quad \zea = 0, \\
	\label{eq:war_zero_lambda_0_02}
	\yea = 0, & \quad \zea = 0, \\
	\label{eq:war_zero_lambda_0_03}
	\xej = 0, & \quad \zej = 0.
\end{align}
It is important to emphasize that these lines cannot be seen as equilibrium manifolds and the tracking error $\m{e^a}$ does not converge to them.

Now, let us investigate the boundedness of the control input vector $\m{\tau}$. Recalling \eqref{eq:control:g_disturbance}, we can show that function \(\m{g}\) is bounded when velocities of the object and its desired trajectory, as well as the time derivative of the desired trajectory, are also bounded. Moreover, due to the definition of $\mXi$ given by \eqref{eq:mXi}, the following components of the auxiliary tracking error should satisfy the following inequalities
\begin{align}
	\pe > 0,  \quad
	\pej > 0, \quad
	\ke > 0,  \quad
	\oe > 0.
	\label{eq:stability_conditions_on_errors}
\end{align}
This requirement is met due to the offset $\m{\delta}$ in \eqref{eq:sigma}. Hence, error $\m{e}$ on the sliding surface is pushed away from zero by design.

\section{NUMERICAL VALIDATION} \label{sec:num_valid}

An important feature of the basic sliding mode control is chattering, i.e. the effect of high-frequency switching of the sign of the control signal in the vicinity of the sliding plane. In the validation, this effect is reduced by replacing the signum function with the following sigmoid function:
\begin{align}
    \label{eq:funkcja_sigmoidalna}
    \forall \zeta\in\mathbb{R},\, \text{sgm}(\zeta) = \left(\abs{\zeta}+ {}^n \epsilon \cdot \abs{{}^n e}\right)^{-1}\zeta, \, n \in \{p, k, o\},
\end{align}
where ${}^n\epsilon>0$.

To verify the proposed control solution, we have conducted extensive simulations. In this section we present simulation results obtained for a simplified model of dynamics in the form:
\begin{align}
    \label{eq:dynamika_uproszczona}
    \dgaaa = \m{M}^{-1}
    \left(\m{\tau}-\m{C} \gaaa - \m{C}_p \diag(\sign(\gaaa)) \left(\gaaa\right)^2\right),
\end{align}
where: $\m{M} = \m{C}_p = \m{1}$ and
\begin{align}
    \m{C} =       & \BM \m{0}     & -\m{S}(\nuaa) \\
    -\m{S}(\nuaa) & -\m{S}(\omaa)                 \\
    \EBM.
\end{align}
The simplified model is used to verify the convergence of the trajectory error to an arbitrarily small level, which is shown by the theoretical analysis.

The control parameters assumed in the simulation are:
\begin{align}
    \m{\delta} = & \left[0.2\ 0.01\ 0.01\right]^\top, \,
    \m{K} = 0.1\times \m{1},                                        \\
    \m{\Rho} =   & 5\times\m{\hat{1}},\, \m{\hat{M}}=10\times\m{M}.
\end{align}
\begin{figure}[!t]
    \centering
    \includegraphics[]{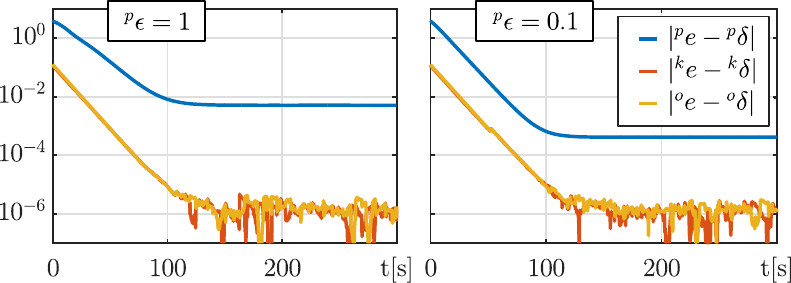}
    \vspace{-0.3cm}
    \caption{The comparison of time evolution of norms of \textsl{error terms} \((\m{e}-\m{\delta})\), where \({}^p \epsilon=1\) and \({}^p \epsilon = 0.1\)}
    \vspace{-0.5cm}
    \label{fig:sym_01AB_smc_e_log_delta}
\end{figure}
Initially, the body frame of the modeled blimp is parallel to the global frame; its position is defined by $\m{p}^b_a = \left[10\ 20\ -30\right]^\top$ m, while the desired trajectory is chosen as $\pbd(t) = \left[0.1 t\ 0 \ 0\right]\T$ m.
Figure \ref{fig:sym_01AB_smc_e_log_delta} shows the comparison of the results of the simulations with parameters \({}^k \epsilon\) and \({}^o \epsilon\) equal to \(1\) for two values of parameter \({}^p \epsilon\in\left\{0.1, 1 \right\}\). It is clear that the decreasing value of ${}^k \epsilon$ improves the control precision on the sliding manifold, namely, the value of \(\abs{{{}^p e}-{{}^p \delta}}\) becomes lower in the steady state. These results are consistent with the theoretical analysis and confirm that $\sigma$ converges to a given neighborhood of zero, whose radius can be made arbitrarily small.

\section{EXPERIMENTAL VERIFICATION} \label{sec:exp_valid}
\begin{figure}[!b]
    \centering
    \vspace{-0.4cm}
    \includegraphics[width=0.45\linewidth]{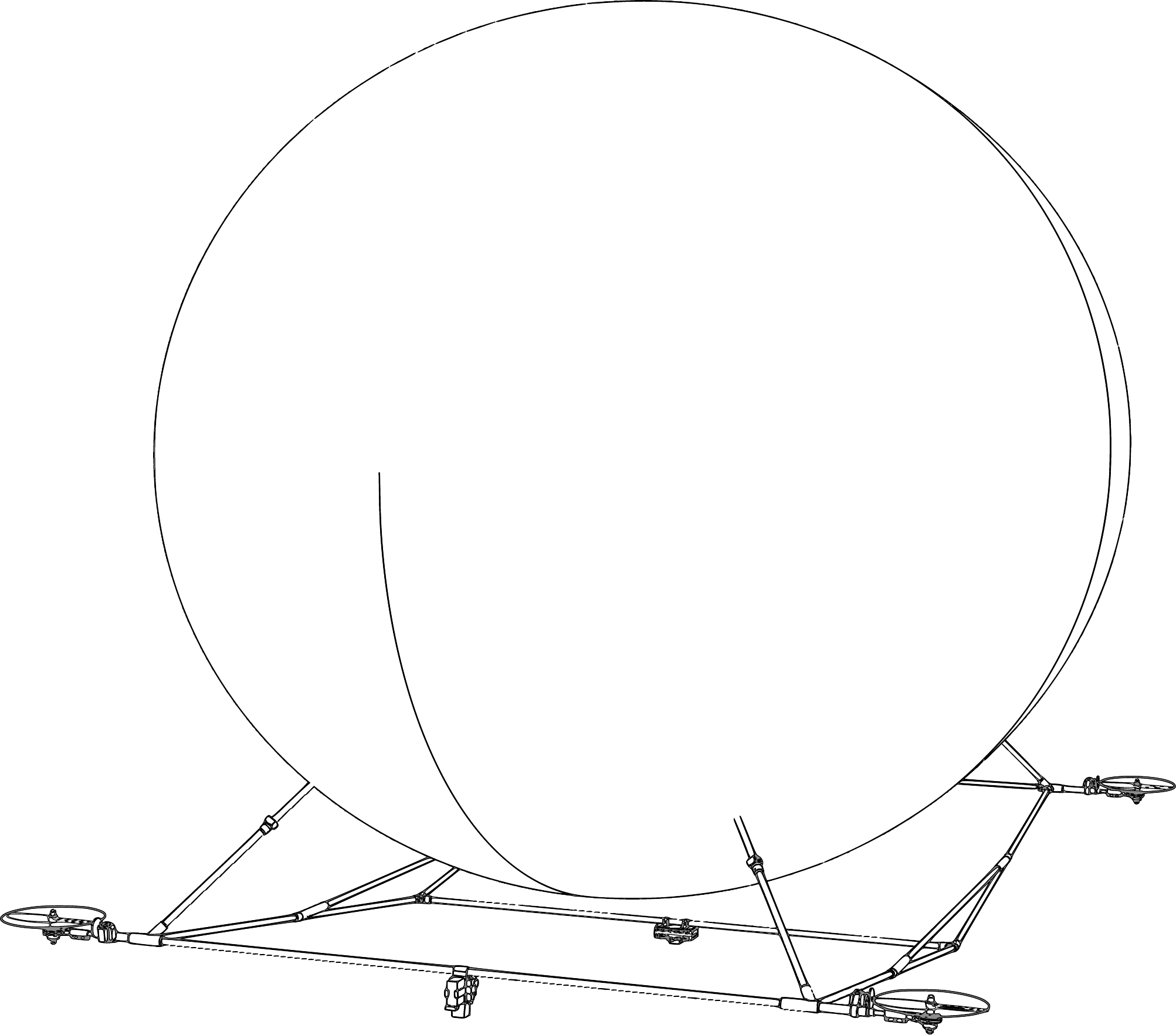}
    \includegraphics[width=0.45\linewidth]{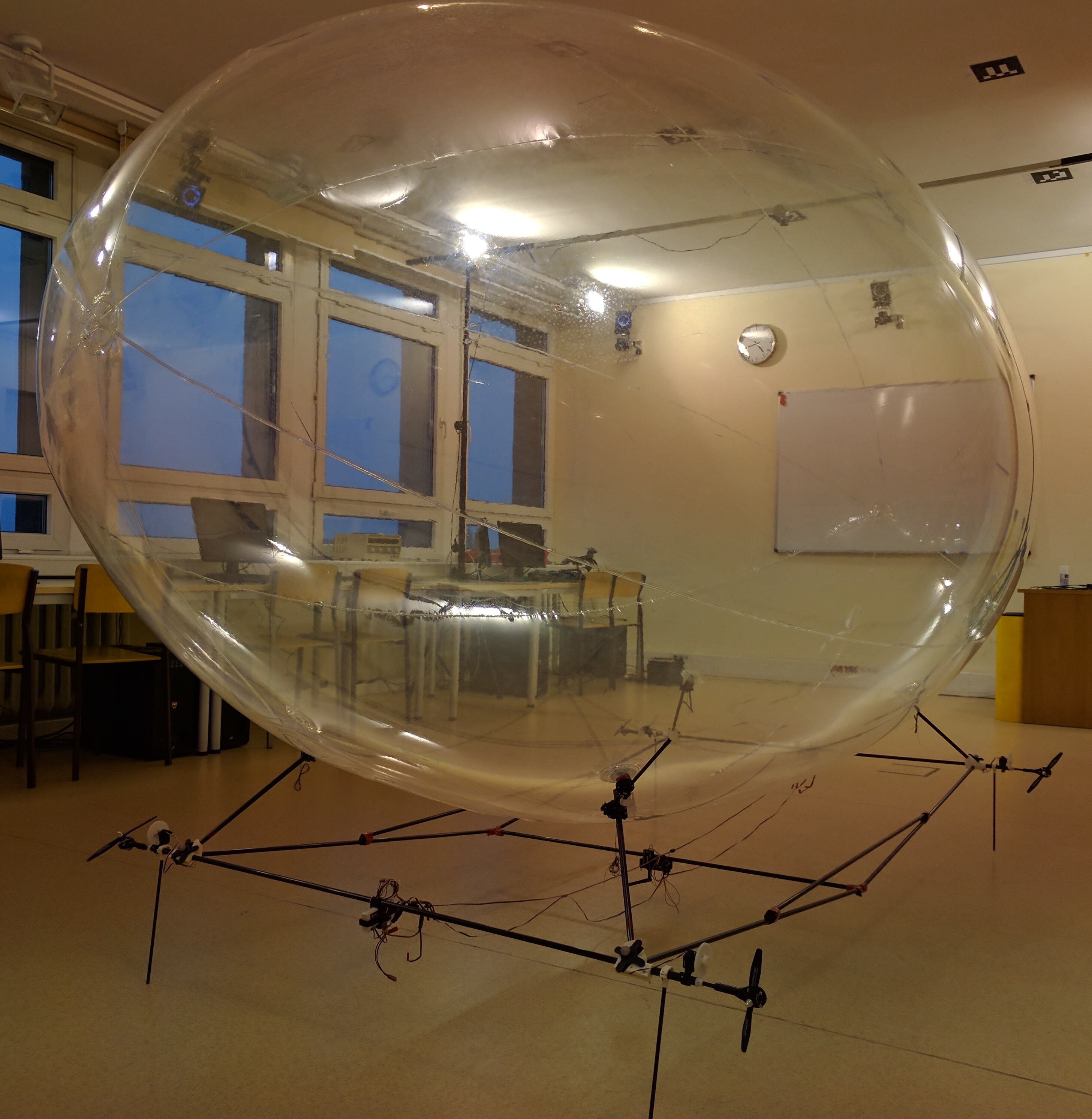}
    \vspace{-0.5em}
    \caption{Sketch of the project and a photograph of the airship}
    \label{fig:sterowiec_projekt_zdjecie}
\end{figure}

\begin{figure}[!b]
    \centering
    \includegraphics[width=0.7\linewidth]{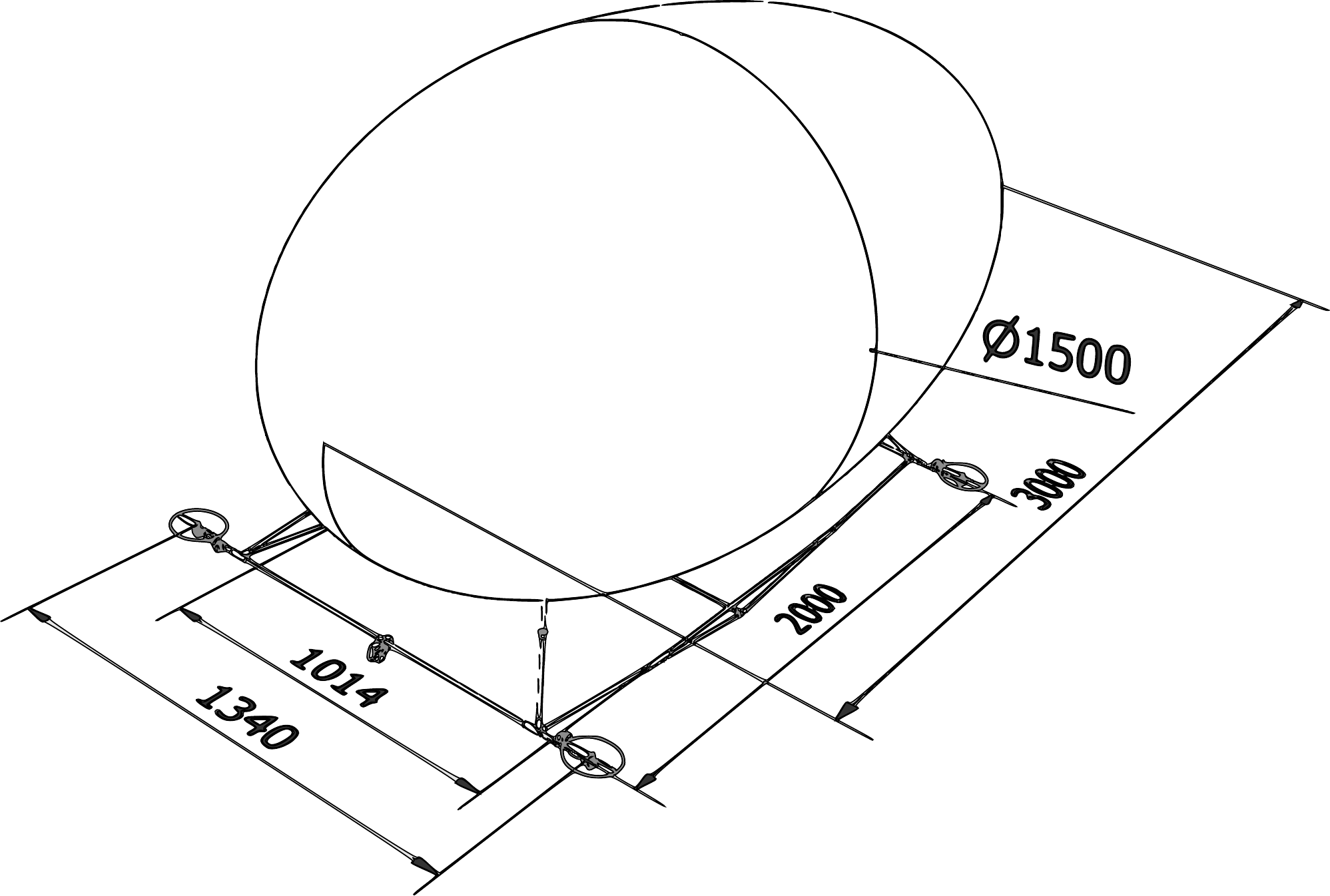}
    \vspace{-0.5em}
    \caption{Model of the airship with dimensions}
    \vspace{-0.5em}
    \label{fig:wymiary_sterowca}
\end{figure}

The environment in which airships operate is particularly difficult to model, so it is extremely important to verify theoretical results in real-life conditions. Therefore, to make it possible to conduct experiments, the authors constructed a test platform in the form of an autonomous airship, (cf. Fig. \ref{fig:sterowiec_projekt_zdjecie}).

The size of the airship (Fig.~\ref{fig:wymiary_sterowca}) has been chosen in such a way that it is possible to mount on it four engines placed in a horizontal plane. Each engine can rotate independently around an axis parallel to the Y-axis of the body frame. This type of propulsion appears to be advantageous because it is possible to use the engine power to compensate for the weight of the object.

\vspace{-0.2cm}
\subsection{The test platform}

The expected weight of the structure without the envelope was about 1.5 kg, taking into account the batteries that allow the operation of the engines for 15 minutes, at full load. Therefore, the dimensions of the balloon were first estimated at 3 m in length, with a maximum diameter of 1.5 m.
The actual weight of the frame before the envelope was mounted on it was 1.65 kg. Therefore, a 3.35 m long polyurethane envelope with a maximum diameter of 1.66 m was made. The volume of the bearing gas was approximately 4.9 m\textsuperscript{3}. The envelope of this size allowed us to obtain a buoyancy compensating the weight of the whole structure, with a reserve of about 0.2 kg.

\begin{figure}[!t]
    \centering
    \includegraphics[]{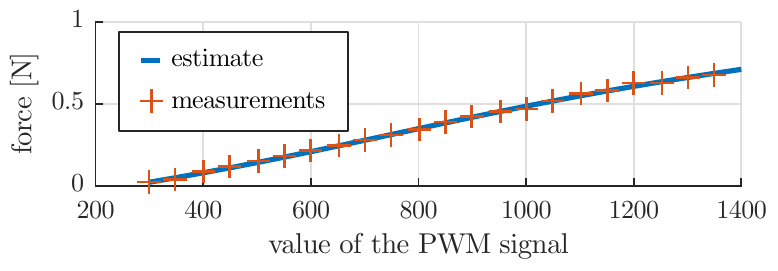}
    \vspace{-0.3cm}
    \caption{The experimentally collected and estimated thrust characteristics of the engines}
    \vspace{-0.5cm}
    \label{fig:sila_silnikow}
\end{figure}

The engines were controlled in an open loop, based on their estimated characteristics (Fig.~\ref{fig:sila_silnikow}).

In the experiments, an external visual localization system was used.
In order to check the robustness of the proposed solution with respect to inevitable measurement errors, in Experiment 2 presented below the measurements were intentionally deteriorated.

A desktop computer communicating with the airship by radio modules with sampling frequency exceeding 100Hz served as a main control unit.

\vspace{-0.2cm}
\subsection{Experiments}

Similarly as in the simulation part, the signum function was replaced by a sigmoidal function (\ref{eq:funkcja_sigmoidalna}). Parameters of the controller were selected as follows: \({}^n \epsilon = 0.1, \, n \in \{p, k, o\}\),
\begin{small}
    \begin{align}
        \m{K} =    & \BM 0.1              & 0           & 0                                          \\
        0          & 0.2                  & 0                                                        \\
        0          & 0                    & 0.2\EBM, \,
        \m{\delta} = \BM 0.2                                                                         \\ 0.01 \\ 0.01 \EBM,\, \m{\hat{M}} = 10\times \m{1},\\
        \m{\Rho} = & \diag \left(\BM 0.06 & 0           & 0.015 & 0.003 & 0.003 & 0.03 \EBM \right).
    \end{align}
\end{small}
The control signal \(\m{\tau}\) was realized by the propulsion system using the following transformation:
\begin{footnotesize}
    \begin{align}
        \m{f}_{xz} = & \m{B}_{xz}^\dagger \m{\tau}
        =
        \left[
            {}^{1x} f_{xz}\, {}^{1z} f_{xz}\, {}^{2x} f_{xz}\, {}^{2z} f_{xz}\,
            {}^{3x} f_{xz}\, {}^{3z} f_{xz}\, {}^{4x} f_{xz}\, {}^{4z} f_{xz}
            \right]\T,                                                                                   \nonumber \\
        \m{B}_{xz} = & \BM
        1            & 0                           & 1      & 0      & 1      & 0      & 1      & 0                \\
        0            & 0                           & 0      & 0      & 0      & 0      & 0      & 0                \\
        0            & 1                           & 0      & 1      & 0      & 1      & 0      & 1                \\
        0            & y_{1}                       & 0      & y_{2}  & 0      & y_{3}  & 0      & y_{4}            \\
        z_{1}        & -x_{1}                      & z_{2}  & -x_{2} & z_{3}  & -x_{3} & z_{4}  & -x_{4}           \\
        -y_{1}       & 0                           & -y_{2} & 0      & -y_{3} & 0      & -y_{4} & 0
        \EBM, \nonumber
    \end{align}
\end{footnotesize}
\vspace{-0.2cm}
\begin{align}
    f_i =      & \sqrt{\norm{\BM {}^{ix} f_{xz}                               & {}^{iz} f_{xz}\EBM}}, \\
    \alpha_i = & \ATAN{ {}^{iz} f_{xz}}{{}^{ix} f_{xz}},\, i \in \{1,2,3,4\},
\end{align}
where \({}^{ix}f_{xz}\), \({}^{iz}f_{xz}\) are the forces that should be generated by the $i^{th}$ engine in the direction of the \(X\)-axis and the \(Z\)-axis of the body frame respectively. Indices \(i \in \{1,2,3,4\}\)denote front-left, front-right, rear-left and rear-right engines, respectively. The~\(\dagger\)~symbol indicates the Moore-Penrose pseudo-inversion.
Parameters \(x_i, y_i, z_i\) specify the distances along the axis \(X, Y, Z\) of subsequent engines from the center of gravity of the object.
The expression \(\ATAN{y}{x}\) denotes a two-argument function of arcus tangens,
while \(f_i\) and \(\alpha_i\) denote, respectively, the force that should be generated by the \(i\)-th engine and the angle at which it should rotate.

\vspace{0.1cm}
The desired trajectory is described by the equation:
\begin{small}
    \begin{align}
        \label{eq:tangensoida_wzor}
        \pbd(t) = \BM
        0.05 \cdot t                                                                                       \\
        0.25 \cdot \tanh(t \cdot 0.075-3)-0.25 \cdot \tanh(-3)                                             \\
        -0.1 \cdot \sin\left(t \cdot 0.0393-\frac{\pi}{2}\right)+0.1 \cdot \sin\left(-\frac{\pi}{2}\right) \\
        \EBM
    \end{align}
\end{small}

\subsubsection*{\textbf{Experiment 1}) The experiment with the best available measurements}

\begin{figure}[!t]
    \centering
    \includegraphics[]{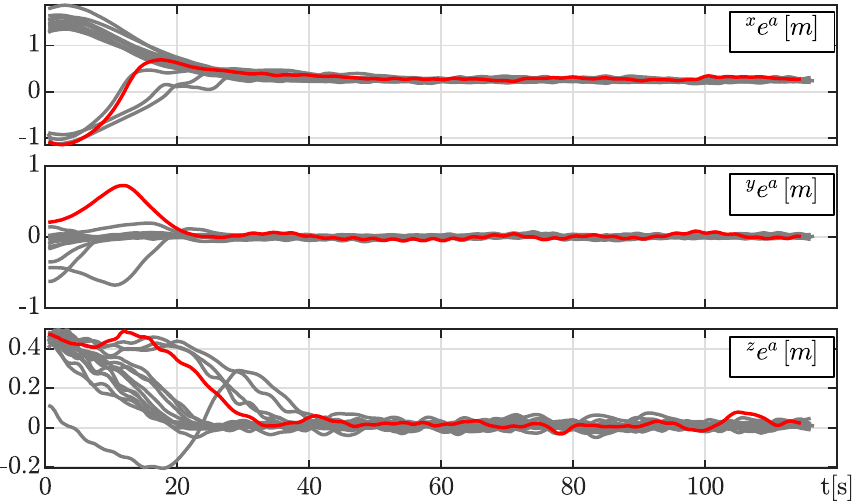}
    \vspace{-0.3cm}
    \caption{Experiment 1: Components of the position tracking error measured in the body frame}
    \vspace{-0.2cm}
    \label{fig:eksp_02_e_g_comp}
\end{figure}
\begin{figure}[!t]
    \centering
    \includegraphics[]{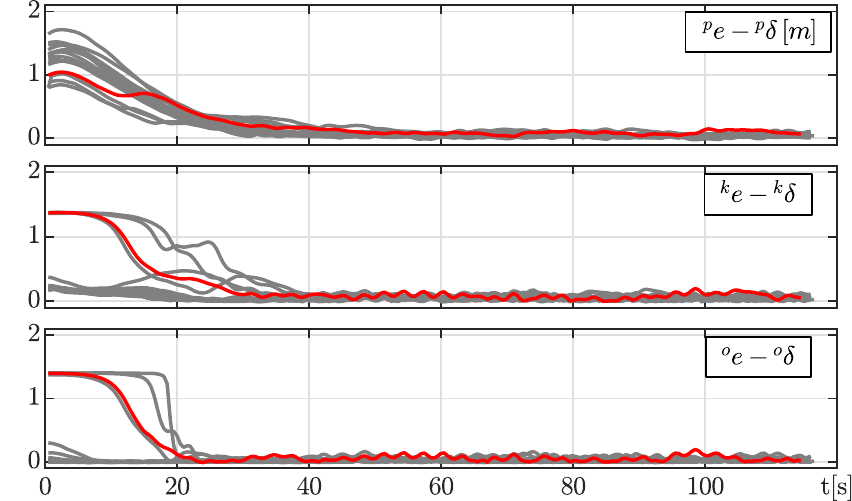}
    \vspace{-0.2cm}
    \caption{Experiment 1: Values of the components of the expression \(\m{e}-\m{\delta}\); note that \(\ke\) and \(\oe\) are unitless quantities}
    \vspace{-0.5cm}
    \label{fig:eksp_02_smc_e_comp}
\end{figure}
\begin{figure}[!t]
    \centering
    \includegraphics[]{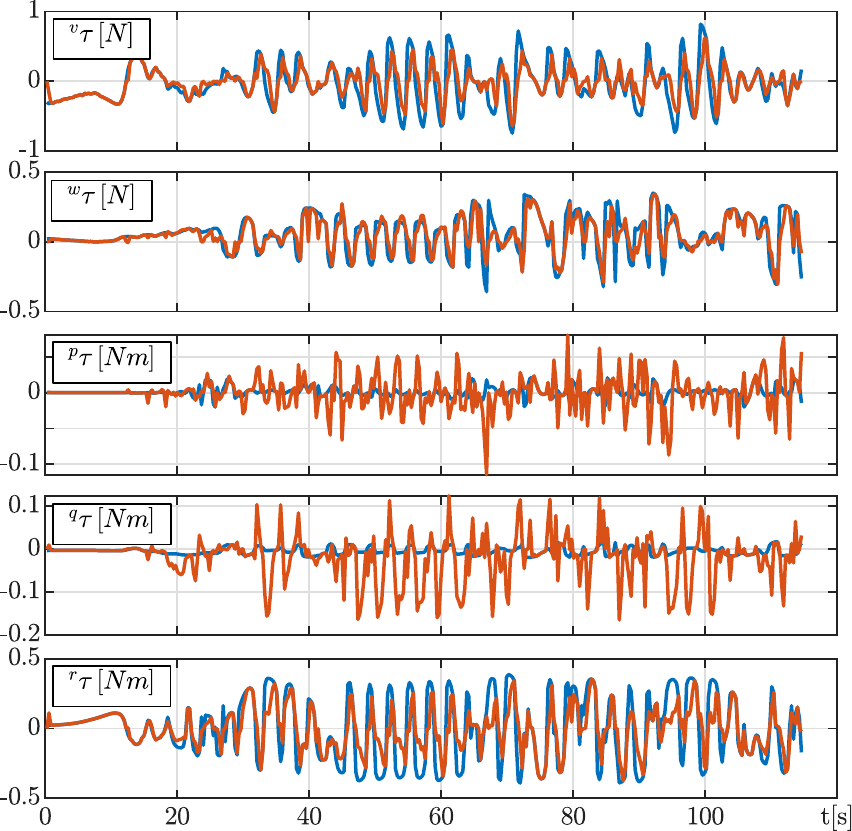}
    \vspace{-0.2cm}
    \caption{Experiment 1: Values of control signals. In blue, the values at the controller output. In orange, the values calculated on the basis of low-level commands which take into account physical limitations of the actuators.}
    \vspace{-0.2cm}
    \label{fig:eksp_02_tau}
\end{figure}
\begin{figure}[!t]
    \centering
    \includegraphics[]{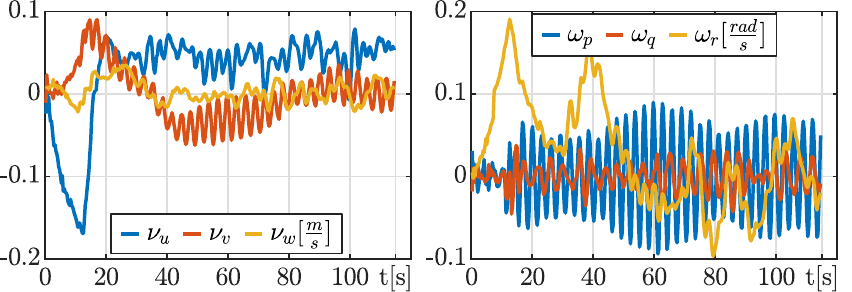}
    \vspace{-0.2cm}
    \caption{Experiment 1: Local velocities of the airship}
    \vspace{-0.2cm}
    \label{fig:eksp_02_nu}
\end{figure}
\begin{figure}[!t]
    \centering
    \includegraphics[]{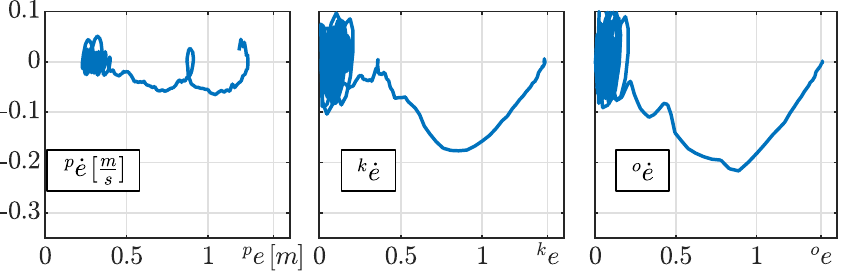}
    \vspace{-0.2cm}
    \caption{Experiment 1: Phase portraits of auxiliary errors; note that \(\ke\) and \(\oe\) are unitless quantities}
    \label{fig:eksp_02_sigma}
\end{figure}
\begin{figure}[!t]
    \centering
    \includegraphics[]{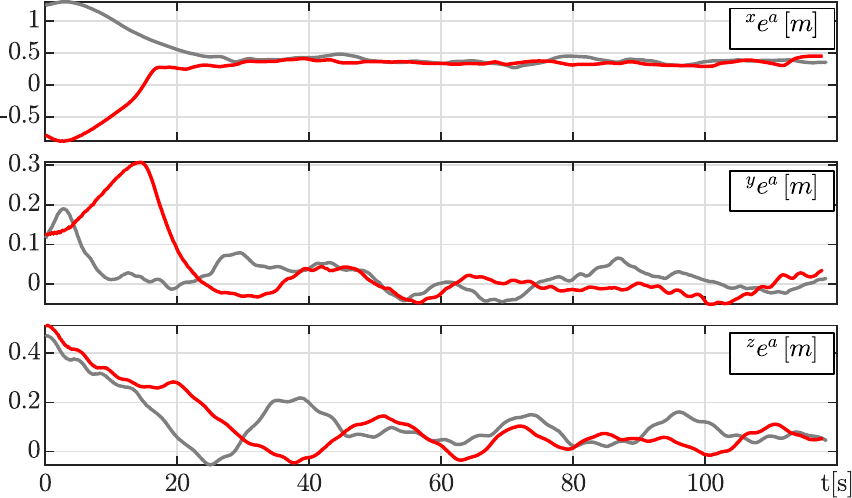}
    \vspace{-0.3cm}
    \caption{Experiment 2: Components of the position tracking error measured in the body frame}
    \vspace{-0.5cm}
    \label{fig:eksp_05_e_g_comp}
\end{figure}
\begin{figure}[!t]
    \centering
    \includegraphics[]{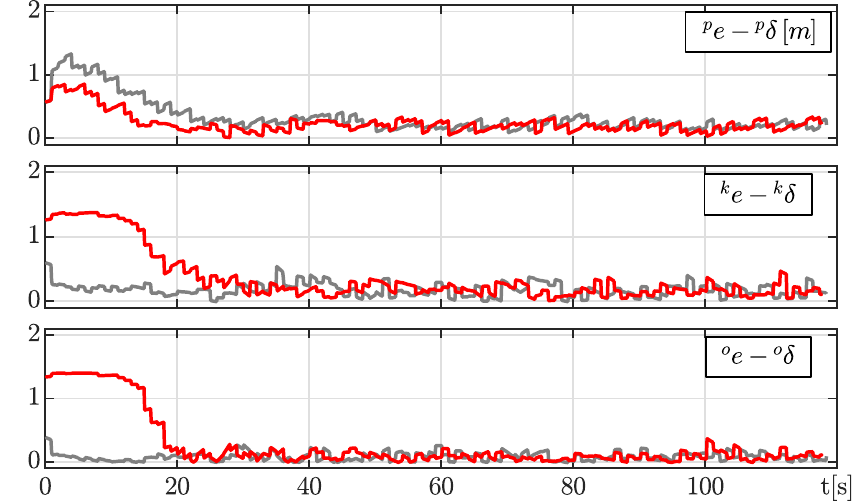}
    \vspace{-0.3cm}
    \caption{Experiment 2: Values of the components of the expression \(\m{e}-\m{\delta}\); note that \(\ke\) and \(\oe\) are unitless quantities}
    \vspace{-0.3cm}
    \label{fig:eksp_05_smc_e_comp}
\end{figure}

Figures \ref{fig:eksp_02_e_g_comp} and \ref{fig:eksp_02_smc_e_comp} illustrate the cumulative results of 15 flights. The red color is used to indicate one selected flight, the detailed results for which are shown in Figs \ref{fig:eksp_02_tau}-\ref{fig:eksp_02_sigma}. The selected flight is also presented in the enclosed video.

The form of local errors (Fig.~\ref{fig:eksp_02_e_g_comp}) and auxiliary errors (Fig.~\ref{fig:eksp_02_smc_e_comp}) shows good repeatability of the proposed algorithm for different initial configurations.

The oscillations in velocities (cf. Fig.~\ref{fig:eksp_02_nu}), especially \(\omega_p\), are caused by the propulsion system delays.
The impact of this effect can also be observed in the phase portraits of auxiliary error (Fig.~\ref{fig:eksp_02_sigma}) in the vicinity of \(\m{\delta}\).

\begin{figure}[!t]
    \centering
    \includegraphics[]{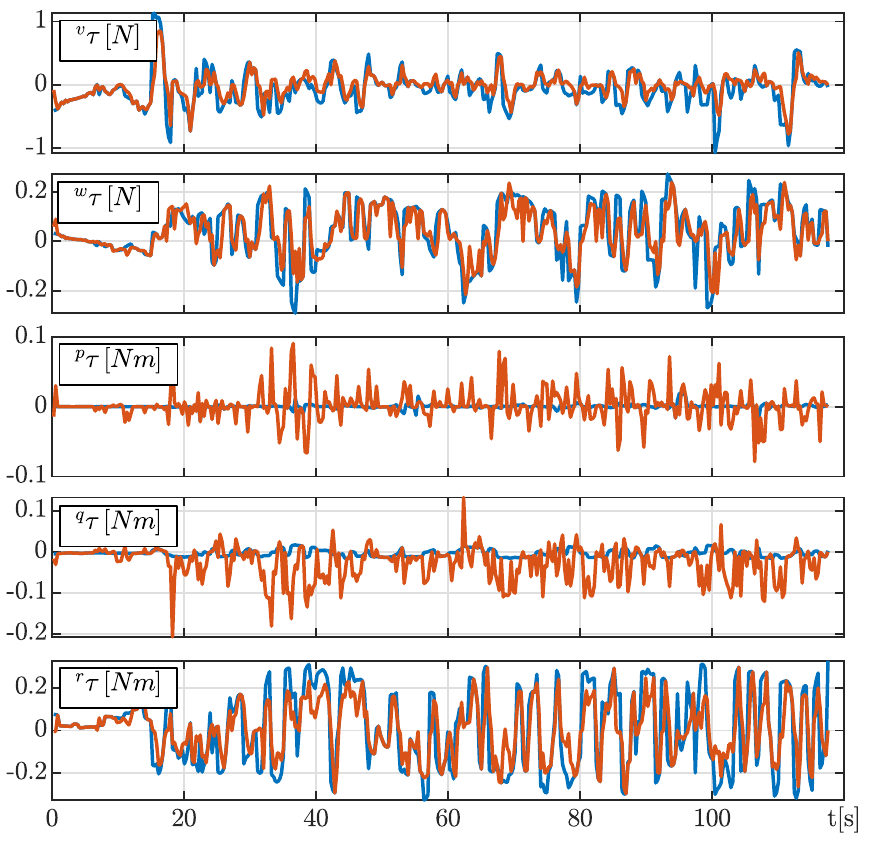}
    \vspace{-0.7cm}
    \caption{Experiment 2: Values of control signals. In blue, the values at the controller output. In orange, the values calculated on the basis of low-level commands which take into account physical limitations of the actuators}
    \vspace{-0.2cm}
    \label{fig:eksp_05_tau}
\end{figure}
\begin{figure}[!t]
    \centering
    \includegraphics[]{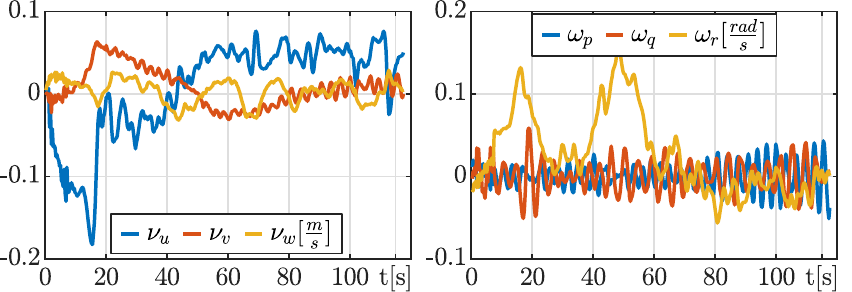}
    \vspace{-0.2cm}
    \caption{Experiment 2: Local velocities of the airship}
    \vspace{-0.2cm}
    \label{fig:eksp_05_nu}
\end{figure}
\begin{figure}[!t]
    \centering
    \includegraphics[]{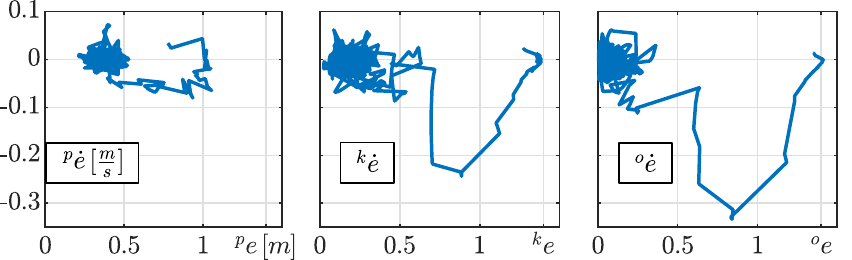}
    \vspace{-0.3cm}
    \caption{Experiment 2: Phase portraits of auxiliary errors; note that \(\ke\) and \(\oe\) are unitless quantities}
    \vspace{-0.3cm}
    \label{fig:eksp_05_sigma}
\end{figure}

\subsubsection*{\textbf{Experiment 2}) The experiment with deliberately deteriorated measurements}

The only difference between Experiments 1 and 2 is the introduction of a pseudo-random disturbance in the range \([-0.2, 0.2]\,m\) in the measurement path and the reduction of its frequency to 1 Hz.
This change was made to check the robustness of the algorithm against measurement errors.
In spite of introducing the disturbances, the local position errors (Fig.~\ref{fig:eksp_05_e_g_comp}) and auxiliary errors  (Fig.~\ref{fig:eksp_05_smc_e_comp}) indicate only a minor decrease in control quality -- cf. Table~\ref{tab:porownanie_bledow_eksp}. As in Experiment 1, the red line indicates the flight whose detailed results are presented in Figs \ref{fig:eksp_05_tau}-\ref{fig:eksp_05_nu}. The selected flight is also presented in the enclosed video.

The phase portraits of auxiliary errors (Fig. \ref{fig:eksp_05_sigma}) deteriorated as a result of step-wise changes in the measured value of the position. However, the disturbances did not cause significant changes in the value of control signals (Fig.~\ref{fig:eksp_05_tau}) or local velocities (Fig.~\ref{fig:eksp_05_nu}).

According to the knowledge of the authors, to date, there have been no results of experimentally verified research that would allow for a reliable comparison of the presented solution.

\begin{table}[!t]
    \caption{Comparison of the position error values expressed in the body frame. The maximum absolute values of the components of the position error and the maximum value of the norm of vector \(\ea\) for \(t > 80\). \vspace{-1em}}
    \label{tab:porownanie_bledow_eksp}
    \begin{center}
        \begin{tabular}{|c||c||c|}
            \hline
            experiment                              & \begin{tabular}{@{}c@{}}best \\ measurements\end{tabular} & \begin{tabular}{@{}c@{}}deteriorated \\ measurements\end{tabular} \\
            \hline
            \(\max{\left(\abs{\xea}\right)}\, [m]\) & 0.3455                     & 0.4558                     \\
            \hline
            \(\max{\left(\abs{\yea}\right)}\, [m]\) & 0.0803                     & 0.0513                     \\
            \hline
            \(\max{\left(\abs{\zea}\right)}\, [m]\) & 0.0797                     & 0.1114                     \\
            \hline
            \(\max\left(\norm{\ea}\right)\, [m]\)   & 0.3497                     & 0.4594                     \\
            \hline
        \end{tabular}
    \end{center}
    \vspace{-1em}
\end{table}

\section{CONCLUSIONS} \label{sec:conclusions}
Currently available sensing technologies allow for relatively accurate measurements of orientation and angular velocity. However, the measurements of linear velocity and the position of an object are still a major problem.
The precision of the measurement system influences the performance of the proposed algorithm,
but its main advantage is the ability to obtain stable operation with relatively low accuracy of this system, which is frequently unachievable in the case of adaptive algorithms.

In the case of airships, precise identification of dynamic parameters of the object is a difficult task. Additionally, changing environmental conditions may influence the values of the parameters. It is a property that significantly complicates the practical use of algorithms based on the model of dynamics. In the proposed solution, the problem of estimation does exist, but possible errors in identifying the parameters do not have a significant influence on the operation of the algorithm. This robustness is the fundamental advantage of the solution.

\bibliographystyle{IEEEtran}
\bibliography{IEEEabrv,Bibliography}
\end{document}